# Phase Separation in Heritage Objects Made of Plasticised PVC—the Case of Joseph Beuys' Multiples

Marwa Saad[a], Sonia Bujok[b*], Aurora Cairoli[c,d,e], Karol Górecki[a,f], Marek Bucki[a,f], Dorota Duraczyńska[b], Dominika Pawcenis[a], Dominika Anioł[g], Kosma Szutkowski[g], Artur Michalak[a], Krzysztof Kruczała[a], Łukasz Bratasz[b]

[a] Faculty of Chemistry, Jagiellonian University in Kraków, Gronostajowa 2, 30-387 Kraków, Poland
[b] Jerzy Haber Institute of Catalysis and Surface Chemistry, Polish Academy of Sciences, Niezapominajek 8, 30–239 Kraków, Poland
[c] University of Milan, Department of Earth Sciences, Via Botticelli 23, 20133 Milan, Italy
[d] Ca' Foscari University of Venice, Department of Environmental Sciences, Informatics and Statistics, Scientific Campus, Via Torino 155, 30172 Mestre (VE), Italy
[e] Sapienza University of Rome, Department of Classics, Piazzale Aldo Moro 5, 00185 Rome, Italy
[f] Doctoral School of Exact and Natural Sciences, Jagiellonian University in Kraków, Łojasiewicza 11, 30-348 Kraków, Poland
[g] NanoBioMedical Centre, Adam Mickiewicz University in Poznań, Wszechnicy Piastowskiej 3, 61-614 Poznań, Poland

*Corresponding author: Sonia Bujok
Email: sonia.bujok@ikifp.edu.pl; postal address: Jerzy Haber Institute of Catalysis and Surface Chemistry, Polish Academy of Sciences, Niezapominajek 8, 30-239 Kraków, Poland

**Abstract**
This work addresses the advanced degradation of plasticised PVC boards, which constitute Joseph Beuys' multiples *Phosphorus–Cross Sled*, focusing on the mechanisms responsible for the extensive exudation of a viscous liquid. A multi-analytical approach combining chromatographic, spectroscopic, microscopic, and mechanical techniques was employed to characterise both the exuded surface layer and the polymer bulk, and to evaluate the consequences of degradation for long-term mechanical behaviour. The experimental results are supported by density functional theory simulations, which indicate a thermodynamic driving force for plasticiser migration towards the PVC surface and its tendency to self-associate. These molecular-level insights provide a mechanistic explanation for the observed migration and exudation phenomena. Overall, the study outlines a coherent deterioration pathway for exudate-covered plasticised PVC. The application of NMR spectroscopy proved to be an efficient method for studying accelerated migration of the plasticiser in PVC, opening the path to the development of non-destructive, *in situ* preventive conservation tools, supporting the early identification of PVC artefacts or parts of a collection that are at risk of severe phase separation.

## 1. Introduction

The deterioration of plastic objects has become a significant challenge for museums and heritage institutions, particularly those with extensive collections of modern art and design[1,2]. Many of these works contain plastic components that were not intended to be long-lasting materials. Until the 1990s, the conservation field paid little attention to the poor ageing properties of plastics, mostly because their instability often remained hidden for decades[3,4]. These degradation phenomena not only compromise the visual qualities of the artworks but also impair their mechanical performance, leading to reduced flexibility and increased stiffness[5,6]. Recent research in this field focuses on the development of standardised approaches for identifying and conserving common types of plastics[7,8] as well as mechanisms of their degradation during accelerated thermal[9,10] and natural ageing[11]. This is another paper from the research carried out within the bilateral project PVCare[12], investigating the degradation of poly(vinyl chloride)—PVC in heritage collections. The already published results focused on the machine learning-supported identification of plasticisers in PVC using spectroscopic techniques[7,8], analysis of the migration of plasticiser caused by the diffusion-evaporation process[6], discolouration (usually yellowing) related to dehydrochlorination of the polymer[13], and emission of acidic degradation products from PVC[14]. Surface stickiness and soiling particle deposition caused by the migration of plasticisers to the PVC surface were also discussed[5]. However, there is still limited quantitative information on the phase separation mechanism leading to the rapid migration of plasticiser, and consequently, PVC degradation within years or decades. The research carried out at the Smithsonian Institution identified that about 1/3 of PVC objects are at risk of fast degradation due to phase separation[6]. This study focuses on Joseph Beuys' *Phosphorus–Cross Sled* (*Phosphor–Kreuzschlitten*) made in the 1970s of plasticised PVC—an extreme case of a phase-separated object[15]. The artwork was donated for research by Schellmann Art Gallery (Munich, Germany), as it cannot be safely handled and exhibited due to liquid exudates covering the surface.

Density functional theory (DFT) is an electronic structure method based on quantum mechanics. It provides information on, i.e., molecular interactions, kinetics, and enables the prediction of macroscopic properties and behaviour of the investigated systems. Even though DFT studies on PVC have recently gained more attention, they mostly focused on thermal degradation and dehydrochlorination of PVC chains[16–19]. On the other hand, DFT calculations on plasticiser–PVC interactions were dedicated to the bio-based and non-phthalate plasticisers[20–23] rather than phthalates present in heritage PVC objects[7,24].

## 2. Research Aim

This research addresses two principal questions: i) why some PVC-based artworks created in the 1970s exhibit severe deterioration while other PVC materials of similar composition and comparable age remain relatively stable, and ii) what molecular mechanisms lead to the formation of the characteristic sticky surface layer observed on these works of art. To develop a detailed understanding of the material deterioration, a broad suite of analytical techniques was employed, including spectroscopic, chromatographic, microscopic, and mechanical methods. In addition, DFT simulations were applied to model the molecular interactions between the plasticiser and the PVC matrix, providing insight into their compatibility and the strength of the interactions. By analysing the associated energy profiles and the plasticiser migration tendencies, the study offers a mechanistic understanding of the processes governing the deterioration of PVC-based artwork.

## 3. Materials and Methods
### 3.1. Materials

All experiments were performed on the pieces of Joseph Beuys' multiples of *Phosphorus–Cross Sled* (*Phosphor–Kreuzschlitten*) artwork (1972–1977, multiple of PVC boards, each of approx. 3 mm thickness).

### 3.2. Methods

All measurements were performed at room temperature unless specified differently for particular experiments or techniques.

#### 3.2.1. Infrared spectroscopy (ATR-FTIR)

The fundamental vibrations and associated rotation-vibration structures were studied in the mid-infrared (IR) region, 4000–550 $cm^{-1}$, using a Thermo Scientific Nicolet 6700 FT-IR spectrometer with ZnSe optics and $LN_2$-cooled MCT detector, equipped with an ATR device (iD5 Diamond Advanced ATR, Thermo Scientific). The ATR-FTIR measurements were performed on various places of the sample with a data interval of 0.482 $cm^{-1}$; 32 scans were averaged for each measurement, and a Happ-Genzel apodization and Mertz phase correction were used.

#### 3.2.2. Nuclear magnetic resonance (NMR) spectroscopy

The liquid NMR measurements were performed using a Jeol JNM-ECZ600R/S1 and Bruker Avance III spectrometers, both operating at 600 MHz. Before the measurements, samples were dissolved in deuterated THF using sonication, resulting in approx. 3 mg $mL^{-1}$ concentration after passing through a 0.45 μm PTFE syringe filter. For the $^1H$ NMR measurements, 16 scans were collected and averaged to create the final spectrum, whereas for the $^{13}C$ NMR measurements, 1024 scans were collected and averaged. Samples were measured threefold—a small sample of the bulk of the artwork was carefully cut, cleaned with isopropanol, and then dissolved; exudate was scraped using a metal spatula from the surface and dissolved (both measured on the Jeol spectrometer—the $^{13}C$ NMR spectrum was predominantly featureless); the surface was swabbed with a cotton swab, which was then put into THF for dissolution of collected exudate (as collected sample was measured on the Bruker spectrometer).

The PGSTE (pulsed-field gradient stimulated spin-echo) NMR experiments were performed on an Agilent DD2 14 T spectrometer (Santa Clara, California, USA), equipped with a Doty Scientific DSI-1374 $^1H$/lock probe head with a maximum magnetic field gradient of 30 T/m. The disc-shaped samples cut from the multiples were placed in the standard 5 mm NMR sample tubes. The stimulated spin echo (STE)[25] was employed to acquire spin echoes. The diffusion time (Δ) was set to 150 milliseconds (with ca 5 ms spin echo time), the magnetic field gradient duration (δ) was set to 2 ms and the magnetic field gradient (*g*) was iterated between 0 and 22 T/m in 32 steps over 32–128 number of accumulations to achieve a sufficient signal-to-noise ratio. Spectra were acquired at 15, 20, 25, and 30 ºC. The data were analysed using MestReNova 14.3, and the diffusion coefficients *D* were fitted using OriginPro 2026.

#### 3.2.3. Electron absorption and reflection (UV-Vis-NIR) spectroscopy and colorimetry

To examine the electronic structure, which affects, among others, the visual appearance of the historic PVC, UV-Vis-NIR spectroscopy was used, and spectra were acquired on an Agilent Technologies Cary 7000 UV-Vis-NIR spectrometer with a halogen (Vis-NIR) and deuterium (UV) lamps as sources and R928 photomultiplier tube (UV-Vis) and cooled PbSmart lead-sulfide-based (NIR) detectors. Measurements were performed in the 300–2500 nm wavelength range (1 nm resolution, 600 nm per minute scan rate, 0.1 s averaging time). The sample was measured both in transmission mode (using air as baseline) and in reflectance mode (in a 150 mm integrating sphere, using Spectralon as reference), after gently cleaning the surface with acetone. For derivative calculations, a 20-point smoothing of the data was performed by the Lowess method beforehand, then the 'derivative' function was used twice in the OriginPro 2025 program. Additionally, colourimetric measurements were performed on a HunterLab ColorQuest XE colour analyser with standard illuminant D65 and standard observer 2°, without specular reflectance, obtaining CIELab parameters.

### 3.2.4. X-ray photoelectron (XPS) spectroscopy

The XPS spectroscopy (R3000 Data Scienta Analyzer-Prevac) was applied to examine the non-cleaned surface composition of an artwork. The base pressure in a vacuum chamber was below $5·10^{-9}$ mBar, and a monochromatized Al-Kα source with a 250 W at 1486.6 eV emission energy was used. The scale of the binding energy value was adjusted to the C 1s reference peak at 284.8 eV. The composition and chemical state of the sample surface were analysed in terms of areas and binding energies of O 1s, C 1s, and Cl 2p photoelectron peaks. The spectra were fitted using CasaXPS software Version 2.3.24PR1.0 (Casa Software Ltd., Teignmouth, UK). The XPS peak fitting was performed using the Shirley background and the GL(30) function. Maximum half-peak widths were limited to 2.0 eV to avoid excessive peaks overlapping and to assign reliable binding energies.

### 3.2.5. Contact angle (CA)

The CA measurements were performed in the atmospheric air. Using the drop shape analysis-profile device with a tiltable plane (DSA-P, Krüss, Germany), contact angles of water and methylene iodide were measured. Further experimental details are available in the Supplementary Materials[26].

### 3.2.6. Gas chromatography-mass spectrometry (GC-MS)

A sample of weight 50 mg was first dissolved in 5 mL tetrahydrofuran (THF, pure for analysis) and agitated for 30 minutes. Then, 10 mL of hexane was added to allow the PVC to precipitate, leaving only extractable additives and degradation products in solution. After filtration, 0.2 mL of phthalate ester standard mixture (7 components) purchased from Sigma Aldrich-Poland (7.5 µg mL$^{-1}$) was added to 0.3 mL of the extracted solution and diluted to a final volume of 1.5 mL using cyclohexane. The analysis was run using an ultra-trace G.C ISQMS (Thermo Fisher-Dreieich, Germany) with a TG5sil/ms capillary column of 30 m length. Helium was the carrier gas with a flow of 1.5 mL min$^{-1}$. The temperature gradient was modified due to decreased flow of carrier gas: 100 °C was stabilised for 1 min and then raised until 320 °C at a rate of 30 °C min$^{-1}$; the final temperature was maintained for a total run-time of 50 min. The solvent delay was 2.5 min. Compounds were identified by their mass spectra and retention indices using the NIST Mass Spectral Library and the Retention Index Database.

### 3.2.7. Size exclusion chromatography (SEC-MALLS-DRI)

Molar mass distributions were determined using SEC chromatography coupled with multiple-angle laser light scattering (MALLS) and differential refractive index (DRI) detectors. The methodology of sample preparation and analysis conditions was described previously[27], and a detailed description is available in the Supplementary Materials.

In the case of liquid exudate, average molar masses were determined based on a conventional calibration approach. SEC setup consisted of the same columns and detectors as above, and the DRI detector was used as the main detector for the calibration with the set of 11 polystyrene standards in the molar mass range 2000–2 000 000 g mol$^{-1}$ (GPC standards, Fluka).

### 3.2.8. Raman microscopy

Raman images were captured using a confocal Raman microscope (WITec alpha300, Ulm, Germany) equipped with a laser operating at 732 nm, connected to the microscope through an optical fibre with a 50 μm diameter.

### 3.2.9. Scanning electron microscopy with energy-dispersive X-ray spectroscopy (SEM-EDX)

To investigate the plasticiser distribution in the artwork, a high-resolution field emission scanning electron microscope, JEOL JSM-7500F (JEOL, Tokyo, Japan), equipped with an energy-dispersive X-ray spectroscopy system, AZtecLiveLite Xplore 30 (Oxford Instruments, High Wycombe, UK), was used for elemental analysis and mapping. Prior to measurements, the sample was coated with a 20 nm layer of chromium using a K575X Turbo Sputter Coater (Emitech Ltd., Kent, UK) to enhance conductivity. SEM-EDX analyses were performed under high vacuum at an acceleration voltage of 15 kV.

### 3.2.10. Uniaxial tensile testing

Owing to the state of degradation and possibility of performing invasive and destructive experiments, dumbbell-shaped specimens (type 1B, half-sized according to ISO 527-3/5 standard) were cut from the boards to perform tensile tests and determine the modulus of elasticity (Young modulus, $E$), strain at break ($\varepsilon$), and tensile strength ($\sigma$). Static tensile tests were performed in a Universal Testing Machine Inspekt Table 10 kN (UTM) from Hegewald & Peschke MPT GmbH (Nossen, Germany) at a loading rate of 500 mm min$^{-1}$. Specimens were machined from the PVC boards of multiples of ca. 3 mm thickness. The $E$ was determined as the slope of the linear fit to the data in the strain range of 0–0.5%.

### 3.2.11. Dynamic mechanical analysis (DMA)

The dynamic mechanical analysis (DMA) was performed using the Triton Tritec 2000 DMA Thermoanalyzer. Frequency sweeps were run in the temperature range of 25–75 °C (with a 10 °C step) at the 0.01–100 Hz frequency range at a constant deformation of 0.002 mm (ca. 0.02% strain) in tension mode to create the master curve by applying the time-temperature superposition principle. Additionally, a temperature scan (DMTA) in the range of 20–90 °C with a heating rate of 2 °C min$^{-1}$ was performed to determine α-transition temperature $T_\alpha$ as the maximum of the tangent of the phase angle $tan\delta$. Specimens of dimensions ca. 1.5x10x20 mm$^3$ were used for all experiments.

### 3.2.12. Density functional theory (DFT) modelling

The DFT calculations were carried out using the Amsterdam Density Functional (ADF) program (version 2017.1 and 2019.3)[28] with the Becke88[29] and Perdew86[30] correlation functional (BP86), and the semiempirical, Grimme D3 dispersion correction with Becke-Johnson damping[31]. For PVC and DOTP, calculations were performed with a triple-ζ Slater basis including two polarisation functions (TZ2P) within the frozen core approximation. For C and O, the 1s core orbitals were kept frozen; for Cl, the 1s, 2s, and 2p orbitals were frozen, while H was treated as a fully free system. Auxiliary s, p, d, f, and g STO functions centred on all nuclei were included to fit the electron density and obtain accurate Coulomb potentials during SCF cycles[32]. Conformational space was explored using the multi-level computational protocol combining DFT calculations with molecular-dynamic simulations on the semi-empirical level of quantum chemistry (PM7), successfully applied for other applications[33–37]. In this approach, semiempirical MD is followed by a set of geometry optimisations for the points from the MD trajectory, first at PM7, and finally at the DFT level. The semiempirical, Born–Oppenheimer molecular dynamics (MD) simulations were performed by a locally developed MD driver program, where the potential energy and the forces acting on nuclei were calculated in single-point PM7[38] calculations. The Verlet-velocity algorithm[39,40] was used for propagating nuclei with a 1 fs timestep. The temperature (fixed at 353 K) in the simulations was controlled by velocity scaling. While the system was warming up, the thermostat was activated at every 5th timestep, and after reaching the desired temperature, at every 150th timestep. For semi-empirical calculations, the MOPAC2016 program was applied[41].

The interaction energy analysis was performed with the extended transition state method combined with natural orbitals for chemical valence (ETS-NOCV)[42–44] as implemented in the ADF program package.

## 4. Results and Discussion
### 4.1. Chemical characterisation and composition

GC-MS was used to identify low-molecular mass components of J. Beuys' artwork (**Fig. S1**). A prominent peak at 3.89 min was observed in the exudate, suggesting the presence of low-molecular-mass components, which agrees with the results by Shin *et al*.[45] for the *Phosphorus–Cross Sled* and *Stamp* sculpture. Another significant peak at 7.38 min retention time was identified as dioctyl terephthalate (DOTP), a commonly used plasticiser in heritage collections[7]. In the case of a 10.67 min retention time, the signal was associated with aromatic compounds, which are formed during PVC degradation as described by Rijavec *et al*.[46]—the first step of dehydrochlorination of pure PVC is polyene formation, followed by the formation of aromatic structures due to molecular cyclisation (backbiting reaction) and chain scission.

SEC-MALLS-DRI analysis was performed to determine the molar masses of the artwork's components, i.e., bulk and surface liquid exudate, as well as the molar mass distribution of PVC. Normalised DRI signal of a bulk sample (**Fig. S2A**, black curve) shows the peak at ca. 19 mL corresponding to PVC macromolecules, whereas other signals starting from ca. 25 mL indicate the presence of lower-molar mass compounds, including DOTP. The liquid exudate contains only lower-molar mass compounds (the largest compound of ca. 1.1-1.2 kg mol$^{-1}$, **Table S1**). The normalised molar mass distributions (**Fig. S2B**) of samples from which the surface exudate layer had been removed ($M_w$ = 112.0 kg mol$^{-1}$) and of non-cleaned samples ($M_w$ = 111.9 kg mol$^{-1}$) are comparable within experimental error (**Table S1**). The measured weight-average molar mass ($M_w$) and dispersity ($Đ$) of the PVC fall within the range of typical $M_w$ (75–186 kg mol$^{-1}$) and $Đ$ (1.4–1.8) values previously reported for PVC in heritage objects[27].

The liquid state $^1$H NMR spectrum of samples taken from the bulk (**Fig. S3A**) revealed resonances characteristic of PVC[47], as well as of the plasticiser and other additives present in the material. These signals of lower-molar mass compounds are also observed in the $^1$H NMR spectrum of the exudate, indicating that plasticiser and other additives are present both within the bulk and on the surface, albeit at differing concentrations, as reflected by variations in the integrated signal intensities[48]. PVC-related signals are absent on the exudate spectrum. Further improvements in sampling technique through swabbing (in comparison to scraping with a metal spatula) enabled more accurate measurement of the exudate spectrum by ensuring no contamination with the bulk (**Fig. S3B–C**) – the most significant new features are two broad signals on the $^1$H NMR spectrum, one weak around 10 ppm and another around 2.7 ppm, which are characteristic of hydrogen bonding in the system. Most likely, the one downfield can be attributed to a carboxylic acid (a peak is also visible in the $^{13}$C NMR around 175 ppm), and the stronger one indicates the presence of an alcohol in the exudate, together indicating the ongoing hydrolysis only on the surface. It is also supported by stronger polar interactions of the uncleaned surface (higher total surface energy and polar ratio; **Table S2**) compared to the bulk[46]. The XPS analysis of the surface (**Fig. S4**, **Table S3**) reveals chemical states consistent with commonly reported literature values for carbon-, oxygen-, and chlorine-containing species. Notably, the Cl 2p region exhibits a doublet, with the Cl 2p$_{3/2}$ component centred at 200.4 eV. This value falls within the reported range for organochlorine and chloride species (198.0–201.0 eV) formed as a result of PVC degradation via the ionic mechanism, which has been demonstrated to occur under museum conditions[13].

The UV-Vis spectrum (**Fig. S5A**) is featureless, with no discernible extrema, reflecting the translucent appearance of the sample. However, the increase in absorbance towards shorter wavelengths, extending to the detector limit, is attributed to the presence of conjugated double bonds formed in the PVC during dehydrochlorination. This process is indicative of polymer degradation during ageing and is responsible for the discolouration of the artwork, manifested by yellowing[13,49]. In the second derivative of the absorbance spectrum (**Fig. S5B**), the final minimum appears at approximately 470 nm, indicating the presence of conjugated polyene sequences of up to eleven double bonds, which absorb light in the visible region[13]. In the NIR area of the spectrum, absorption bands are visible stemming from the electronic structure (closer to the visible range), as well as from the oscillations (overtones and combination bands) of molecules in the sample, e.g., around 1700 nm and above 2200 nm, which are characteristic of the polymer and the plasticiser[7,50]. To better understand the perception of the artwork's colour and calculate the most common descriptors, CIELab values were determined using a colourimeter with 8°/d geometry:

$$L^* = 58.34 \pm 0.01 \quad a^* = -1.30 \pm 0.01 \quad b^* = 20.45 \pm 0.01$$

where $L^*$ denotes lightness, $a^*$ represents the green-red axis, and $b^*$ is the value on the blue-yellow scale. In total, this indicates a moderate opaqueness (a fully transparent object would have a value of $L^*$ equal to 100) and darkening of the sample, as well as its yellowing (positive $b^*$ value). Therefore, the CIELab values provide an optical indicator of the degradation degree, which supports the spectroscopy results discussed above.

The artwork was further examined using infrared spectroscopy. The cleaned surface was measured as the representation of the bulk of the sample, and the exudate was analysed as a liquid. The main differences in IR spectra (**Fig. S6**) are observed in the low wavenumber range, where bands around 650 cm$^{-1}$ (black curve) characteristic of the PVC chain are present and attributed to oscillations of the C-Cl bond (692, 635, and 612 cm$^{-1}$)[48,50–52]. On the contrary, the exudate spectrum below 1000 cm$^{-1}$ is mostly featureless for the exudate, indicating a lack of chlorinated compounds.

### 4.2. Plasticiser distribution

To investigate the distribution of plasticiser within the artwork's boards, Raman profiles and SEM-EDX were performed. **Fig. 1A–B** shows the Raman depth profiles (cross-sections) of the C=O band in two spots, where the yellow colour corresponds to a higher C=O concentration. The lower intensity in the darker region at the bottom of the images indicates a lower concentration of DOTP in the bulk. Thus, the gradient in intensity (marked with arrows) reflects the concentration profile due to migration of DOTP, which suggests that DOTP is depleting from the bulk as it moves towards the surface. As a result, the increasing C=O intensity in the brighter (yellow) areas at the top suggests a higher concentration of DOTP near the object's surface.

**S**urface mapping of C=O (**Fig. 1C**) shows bright yellow areas representing DOTP agglomeration on the uncleaned artwork's surface, while the dark spots correspond to PVC, suggesting areas where DOTP is absent or less concentrated. This indicates that DOTP has migrated and accumulated, forming distinct clusters or patches. According to Burns *et al.*[53], a clear distinction between bright (DOTP) and dark (PVC) regions supports the hypothesis of phase separation, which occurs due to the incompatibility between the plasticiser and polymer matrix[53,54].

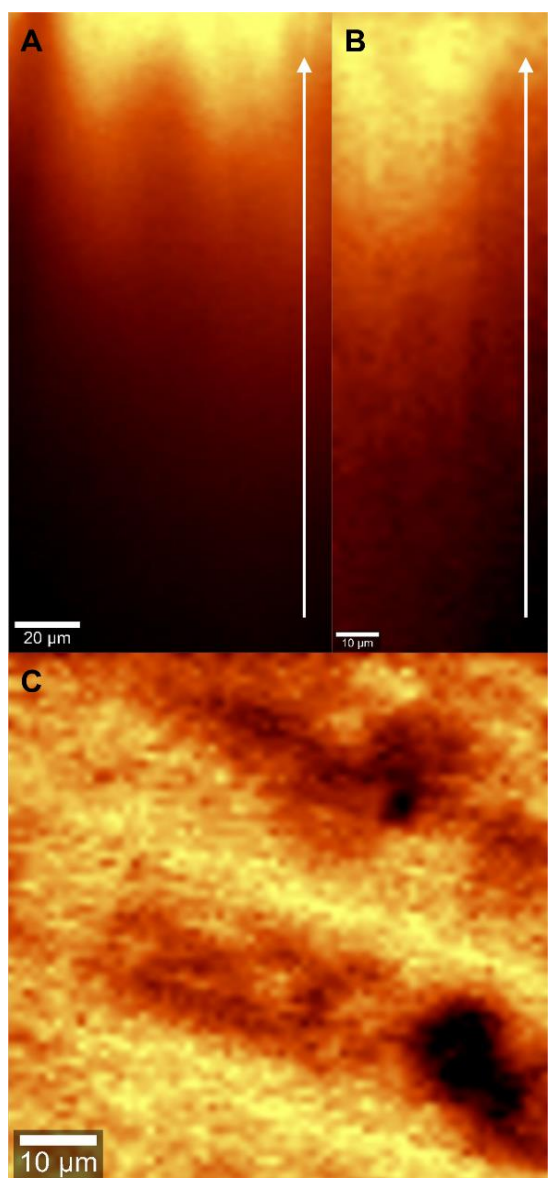

**Fig. 1.** The Raman profile of the C=O band in the sample's cross-section (two spots—**A** and **B**), as well as the artwork's surface (**C**). Yellow colour indicates a higher contribution of the C=O band, thus a larger content of DOTP plasticiser. White arrows in **A** and **B** indicate the DOTP migration path from the bulk towards the surface.

The distribution of DOTP in the PVC matrix was further investigated using SEM (**Fig. 2**) with elemental analysis (EDX) of oxygen and chlorine (**Fig. S7**) in selected cross-section areas: both edges and bulk. It was assumed that oxygen is a representative element of DOTP, which is a major additive, whereas chlorine was attributed to PVC macromolecules. Thus, the higher O/Cl ratio informs us about the potential areas richer in DOTP. First, darker areas close to both edges (top and bottom surfaces of the artwork's boards) were distinguished, and EDX analysis was performed in the rectangular areas marked in **Fig. 2A** and **C**. The resulting O/Cl mass ratio was 0.49 and 0.43 for the top (**A**) and bottom (**C**) edges, respectively. Then, elemental analysis of the bulk area (**Fig. 2B**; entire area) was performed and resulted in the O/Cl mass ratio of 0.30, which is significantly lower and corresponds well with the Raman C=O profiles, indicating DOTP migration towards the board's surface. Moreover, the estimated thickness of the DOTP-rich phase on both edges is ca. 15-20 μm as determined by Raman mapping and SEM (**Fig. 1A–B**; **Fig. 2A** and **C**). Overall, the DOTP-rich phase constitutes around 1.3% of the entire board thickness (40 μm of the total 3 mm).

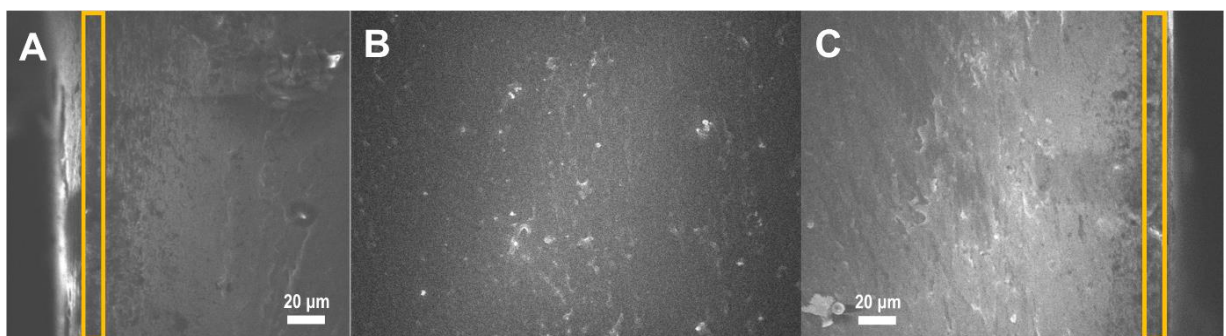

**Fig. 2.** SEM microphotographs of the board's cross-section: **A**—edge close to the top surface, **B**—bulk, and **C**—edge close to the bottom surface.

### 4.3. Mechanical characterisation

Further investigation focused on the effect of phase separation (confirmed by chemical analyses) on the mechanical response of the severely degraded PVC boards (**Fig. S8**, **Table S4**), which has not been studied so far[15,55]. The average value of $E$ was $(236 \pm 13)$ MPa, whereas the average $\varepsilon$ and $\sigma$ values were $(191 \pm 1)$ % and $(16.5 \pm 0.5)$ MPa, respectively, indicating the mechanical behaviour of the boards as ductile material, even though a vast amount of plasticiser exudated. These tensile properties are similar to $\varepsilon$ (216–235%) and $\sigma$ (17–19 MPa) values measured after ca. 25 years of usage of insulation materials made of plasticised PVC (17% m/m of dioctyl phthalate plasticiser) in the early 1970s[56].

In order to estimate the main α-transition temperature $T_\alpha$ (corresponding to the transition from glassy to rubbery state), temperature scans (DMTA) were measured twice (**Fig. S9**). The temperature dependence of $tan\delta$ in the 20–90 °C range revealed the main α-transition in a range of 48–56 °C. However, a small hump on the DMTA curves around 35 °C suggests a different thermo-mechanical response of some part of the board's material. It could be attributed to the DOTP-rich phase (closer to the edges of the boards), whereas the higher (main $T_\alpha$) to the bulk of the boards, which were found to contain less DOTP (see Raman and SEM-EDX results in **4.2**). Herein, it is worth noting that the overall contribution of the DOTP-rich phase is significantly smaller (ca. 1.3% of the total board thickness); thus, all mechanical properties describe the averaged macroscopic behaviour of the PVC boards, primarily attributed to their bulk composition. The average $T_\alpha$ of the main transition (52 °C) was used to estimate the DOTP content of around 15% m/m that remains in the object[6].

The short-term tensile tests provide information about the mechanical behaviour of the material subjected to fast loading rates and deformations. However, in the context of preservation of cultural heritage, long-term deformations are also of practical importance. To investigate the time-dependent mechanical behaviour, the DMA was applied to obtain the master curve (**Fig. 3**), i.e. the long-term dependence of the mechanical response on the frequency (rate) of the applied non-destructive tensile deformation. Generally, DMA provides the storage $E'$ and loss $E''$ moduli corresponding to the elastic and viscous contributions of the polymer material, respectively. To reach the low frequency range that reflects long-term deformations that are inaccessible at the typical laboratory time scale (see top x-axis in **Fig. 3**), the time-temperature superposition principle is applied[57,58]. Briefly, higher temperatures enable faster molecular motion that normally would require a much longer time; thus, at elevated temperatures, such rearrangements of macromolecules occur at a much shorter time scale. This procedure enables the shift of data recorded at higher temperatures towards lower frequencies, reaching e.g. $10^{-10}$ Hz that corresponds to a deformation lasting 300 years, which delivers information on how the object responds to slow and long-term gravitational loads during storage.

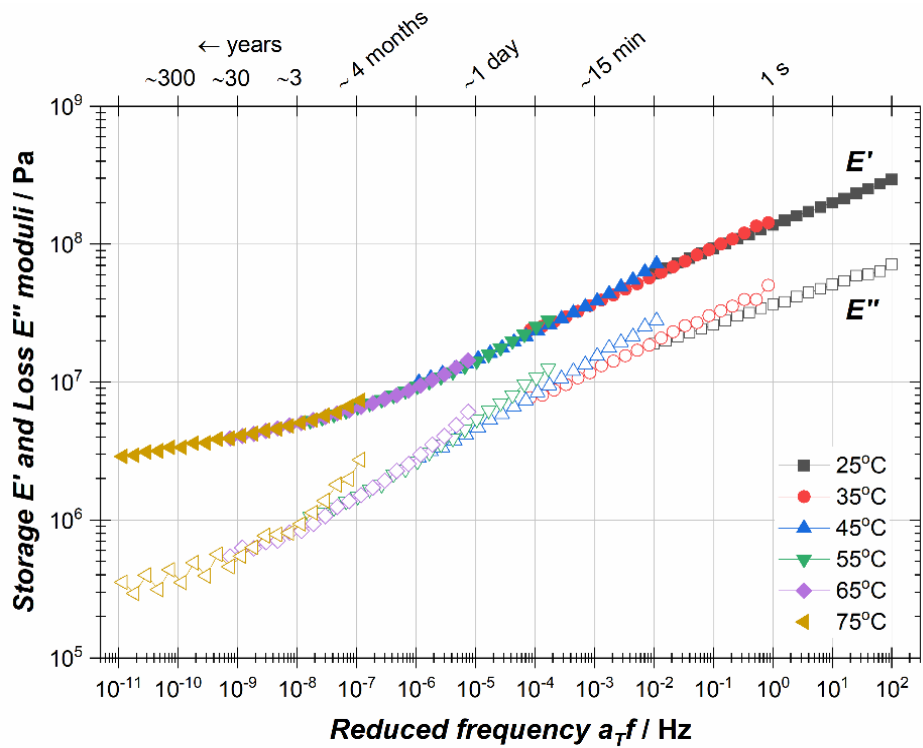

**Fig. 3.** Master curve (time-dependence of storage $E'$ and loss $E''$ moduli) of J. Beuys' artwork at 25 °C.

Master curve (**Fig. 3**) indicates that the material is significantly stiffer ($E'$ of hundreds of MPa) when subjected to sudden impact, e.g., during handling or transportation, which is indicated by higher moduli at a frequency range corresponding to seconds. On the contrary, when permanent or slowly applied loads are considered, moduli drop by two orders of magnitude, which indicates that at this timescale, the material responds similarly to typical elastomers[59].

While creating the master curve, data obtained for each temperature are shifted along the x-axis by known horizontal shift factors $a_T$. Plotting these $a_T$ values as a function of temperature in the Arrhenius-type relation (**Fig. S10**), enables estimation of the activation energy $E_a$ of thermal transition observed within the investigated temperature range. In the case of the investigated sample, the main α-transition was observed; thus, $E_a$ of transition from glassy to rubbery state based on Arrhenius plot (**Fig. S10**) was estimated as $(351 \pm 6)$ kJ mol$^{-1}$, which is a typical value for α-transition in synthetic polymers[60].

### 4.4. Diffusivity of DOTP determined with PGSTE NMR

NMR experiments in the solid state were performed to investigate the diffusivity of DOTP in Beuys' artworks and to compare the effective diffusion coefficient $D$ with available literature data. The obtained signal attenuation curves (**Fig. S11**) indicate the presence of multiple molecular populations characterised by different DOTP mobilities. Therefore, the data were fitted using a triexponential model (**Eq. 1**):

$$I(b) = \sum_i^3 f_i \exp(-bD_i) \quad \text{(Eq. 1)}$$

where $D_i$ is the diffusion coefficient corresponding to the $i$-th distinguished fraction [m$^2$ s$^{-1}$] and $f_i$ is a relative fraction of each population ($\Sigma f_i = 1$). This model enabled the determination of three $D$ regimes at selected temperatures (15, 20, 25, and 30 °C). In particular, at 20 °C $D$ values were: $D_1 = 1.4 \cdot 10^{-12}$ m$^2$ s$^{-1}$, $D_2 = 4.1 \cdot 10^{-13}$ m$^2$ s$^{-1}$, and $D_3 = 4.6 \cdot 10^{-14}$ m$^2$ s$^{-1}$, reflecting at least three differently plasticised phases in the artwork. It also indicates that NMR is the most sensitive of all techniques used in this study for differentiating present areas. With increasing temperature, a systematic increase in all $D$ was observed (**Fig. S12**). For each phase, the activation energy $E_a$ of diffusion was determined as $E_{a1} = 42 \pm 14$ kJ mol$^-$

$^1$, $E_{a2} = 33 \pm 19$ kJ mol$^{-1}$, and $E_{a3} = 82 \pm 28$ kJ mol$^{-1}$. The substantial standard deviations may be attributed to the intrinsic heterogeneity of the degraded polymer system. Variations in local plasticiser concentration may result in a broad distribution of molecular mobilities that non-uniformly respond to temperature changes, leading to deviations from ideal Arrhenius behaviour. According to the literature[61], the $E_a$ of DOTP diffusion in glassy PVC ranges from 57–69 kJ mol$^{-1}$, which is similar to that for DEHP (*ortho-* analogue of DOTP), i.e., 69 kJ mol$^{-1}$, as determined previously[6]. Thus, the highest value of $E_{a3}$ indicates that it can be attributed to DOTP diffusion in the least plasticised phase, with the lowest $D_3$, being in the glassy state—bulk of the artwork. On the contrary, $D_1$ and $D_2$, which are larger by two and one orders of magnitude than $D_3$, respectively, correspond to DOTP-rich areas, with possibly high enough DOTP content to behave more like rubbery, as indicated by lower $E_{a1}$ and $E_{a2}$.

Surprisingly, a comparison of $D_1$ and $D_2$ (**Fig. S12**: dark grey squares and grey circles) with previously obtained correlation[6] (**Fig. S12**: greenish curve $lnD_{rubbery}$) for highly plasticised PVC (rubbery state) indicates that $D_1$ and $D_2$ are larger than the typical diffusivity of plasticiser in rubbery PVC (6.8·10$^{-14}$ m$^2$ s$^{-1}$ at 20 °C) by two and one orders of magnitude, respectively. Moreover, a similar trend for the third diffusivity regime (**Fig. S12**: blue triangles), reflecting a glassy state, was observed (**Fig. S12**: purple curve $lnD_{glassy}$). Previously reported[61] values of $D$ for DOTP in glassy PVC objects (2.3·10$^{-17}$–1.1·10$^{-15}$ m$^2$ s$^{-1}$ at 20 °C) are also 10 to 1000 times lower than $D_3$.

Overall, the obtained NMR results indicate two undeniable proofs of phase separation at the molecular level: i) the distribution of $D$ indicates that the investigated artwork underwent phase separation rather than simple diffusion-evaporation—in the case of the latter, solely single $D$ is observed[5] and ii) diffusivity of plasticiser was significantly larger within all areas in comparison to typical diffusivity in PVC.

The other indication of phase separation is the rate of this process, which is governed by diffusion (concentration gradient) and hydrodynamic effect (interfacial tension gradient)[62]. Assuming that the artwork lost half of the initial DOTP, a purely concentration gradient-driven process (diffusion-evaporation) would have to last around 1200 years to reach the current DOTP content based on the model proposed elsewhere[6], whereas the first signs of phase separation were observed in the mid-90s, 20-25 years after object execution[15]. In the case of purely hydrodynamically driven phase separation, even extremely small changes in interfacial tension would result in significantly faster process, e.g., a 10$^{-6}$% change of interfacial tension would lead to ca. 4 nm s$^{-1}$ flow rate according to the simplified Marangoni effect (estimated based on surface tension and viscosity of DOTP[63])—for a 3 mm thick board, the process would take a few days. Though the timescale of observed phase separation is between these two extremes, this case study suggests that the hydrodynamic effect plays a key role in extensive DOTP exudation within the observed timescale.

From a practical point of view, the temperature dependencies of effective diffusion coefficients determined in this study suggest that each 10-degree temperature drop would slow down diffusivity by approx. 2 times. Although such preventive conservation action is less effective than the typical rule for degradation of organic materials proposed by Stefan Michalski ('Double the life for each five-degree drop')[64], it is the first scientific evidence supporting decisions regarding long-term storage of phase-separating PVC artefacts. Nonetheless, this phenomenon requires special attention and quantification involving both museum professionals and heritage scientists.

### 4.5. DFT modelling of the PVC–DOTP interactions

During experimental investigations in the previous sections, phase separation between PVC and DOTP was confirmed. A computational study was performed to rationalise this behaviour in terms of interactions between DOTP and PVC at the molecular level. In particular, the considered models were constructed to compare the stability and the interactions in the systems in which i) the DOTP molecule is adsorbed on the surface of PVC, ii) the DOTP molecule is embedded in PVC, and iii) the DOTP is surrounded by other DOTP molecules (i.e., as in the plasticiser cavities[5]). Details of the construction of the models (**Fig. S13**) and sampling of their conformational spaces are discussed in the Supplementary Materials.

The minimum-energy geometries for the three cases considered are shown in **Fig. 4**. The relative energies (with respect of the 'global minimum') for sets of geometries modelling interaction of DOTP with PVC are plotted in **Fig. 5**. The results show that, for all configurations, the total energy of DOTP is consistently lower when it is on the PVC surface (black squares in **Fig. 5**) compared to when it is embedded in PVC (red circles in **Fig. 5**). This indicates that DOTP is energetically more stable on the PVC surface than in the bulk. This behaviour aligns with the tendency of DOTP molecules to migrate toward lower-energy configurations, offering a clear explanation for the extended migration of plasticisers, such as DOTP, from PVC objects to their surfaces over time, which is relevant for one-third of PVC collections[6].

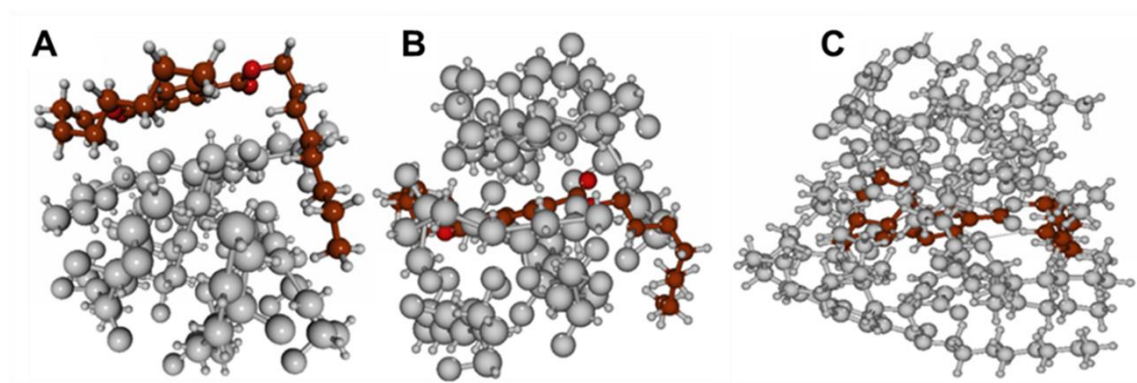

**Fig. 4.** The DFT-minimum-energy geometries of the models with DOTP: **A**—on the PVC surface, **B**—embedded in the PVC, **C**—surrounded by DOTP. For clarity, one DOTP is shown in colour.

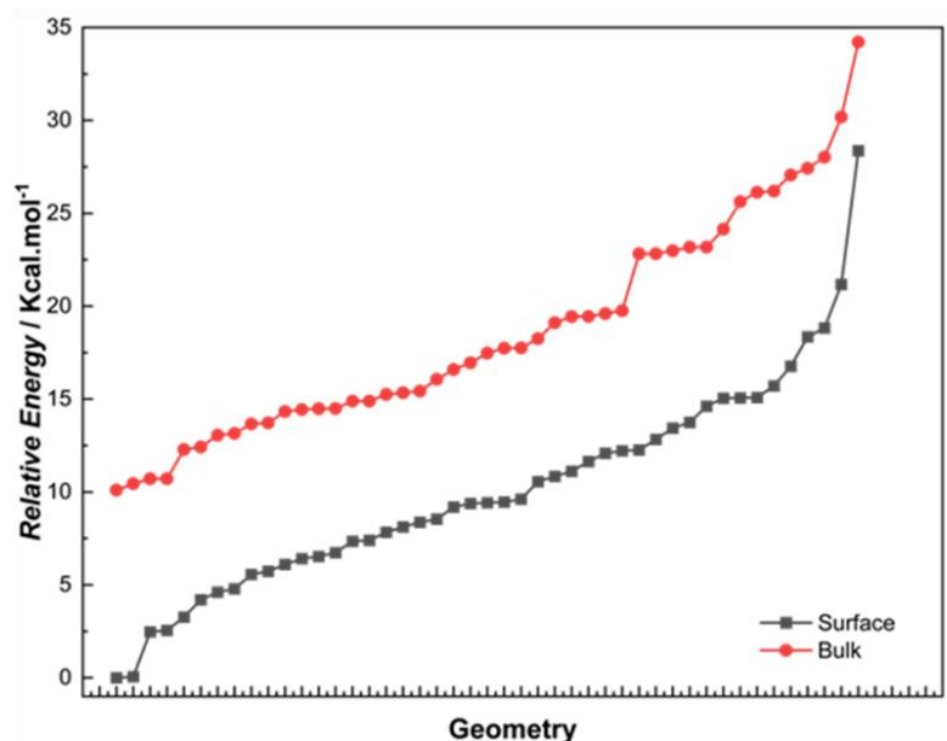

**Fig. 5.** Relative energy of DOTP on the PVC surface (black squares) and in the PVC cluster (red circles) for 50 geometries optimised at the DFT level.

To evaluate the influence of the methodology on the results, the energy difference between DOTP molecules located on the PVC surface and DOTP embedded in the PVC was computed using various DFT functionals (**Fig. S14**). Further, we have evaluated the influence of entropic factors by performing vibrational analysis and computing free energies (**Fig. S15**). The results (**Fig. S14** and **Fig. S15**) consistently show that DOTP exhibits higher stability (i.e., a lower energy) when located at the surface, regardless of the functional employed, thereby reinforcing the robustness of the conclusion.

In the case of the considered models containing the DOTP molecule on the PVC surface and embedded in PVC, a direct comparison of their stability based on the total energies is possible, since they have the same composition (one DOTP molecule and six PVC chains containing 5 monomeric units). Such a comparison of stability based on total energies is not possible for the systems modelling DOTP 'inside' the PVC and 'inside' DOTP clusters. Instead, the energy of the interaction between one DOTP molecule and the rest of the system (PVC or DOTP cluster) can be compared, with the assumption that a stronger interaction leads to a higher stability. Therefore, the Energy Decomposition Analysis combined with Natural Orbitals for Chemical Valence (ETC-NOCV) was used to provide insights into the nature of intermolecular interactions between DOTP within the PVC cluster and DOTP within the DOTP cluster (**Fig. 6**, **Table S5**). The ETS-NOCV analysis was performed with one DOTP molecule considered as one fragment, and the rest of the system as the other fragment. The resulting total bonding energy and its dispersion component are presented in **Fig. 6.** All the bonding energy components resulting from ETS-NOCV analysis are listed in Supplementary Materials (**Table S5**).

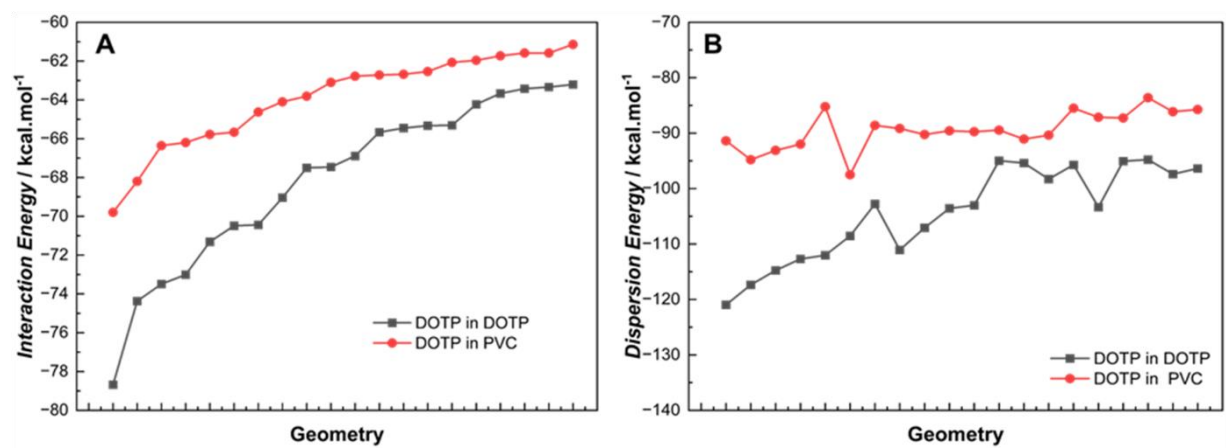

**Fig. 6.** Interaction energy (**A**) and dispersion energy (**B**) of DOTP in DOTP cluster (black squares) and in PVC cluster (red circles) for 20 geometries from DFT calculations. For clarity, only 20 values are plotted in each case, corresponding to the strongest interaction.

The results show that the interaction energy between DOTP molecules (DOTP–DOTP in **Fig. 6A**) is stronger (more negative interaction energy) than that between DOTP and PVC (DOTP–PVC in **Fig. 6A**) for all considered geometries. Further, the results of the interaction energy decomposition analysis (ETS-NOCV) indicate that the difference in the dispersion energy (**Fig. 6B**) is mostly responsible for this effect**.** It is also worth mentioning that further decomposition of the orbital-interaction component into NOCV contributions indicates that they describe mostly polarisation of the fragments, i.e., there are no specific, important non-covalent interactions. In summary, the results indicate a pronounced tendency for the plasticiser to self-associate, with dispersion energy providing greater stabilisation for DOTP molecules surrounded by other DOTP molecules than for DOTP surrounded by PVC, across all geometries.

## 5. Conclusion

This study presents a comprehensive analysis of the degradation mechanisms affecting PVC-based artworks, using Joseph Beuys' *Phosphorus–Cross Sled* as a case study. The results demonstrate that the migration of the DOTP plasticiser from the PVC bulk to the surface is governed by a combination of thermodynamic and hydrodynamic factors. Raman, SEM-EDX, and NMR analyses revealed the formation of distinct DOTP-rich surface domains, consistent with the phase separation concept and responsible for surface tackiness and liquid exudation. The chemical degradation of the PVC matrix, dominated by dehydrochlorination with the formation of conjugated polyene sequences and, at more advanced stages, aromatic compounds, was confirmed by IR, UV–Vis, NMR, and XPS analyses.

Despite substantial plasticiser loss, the bulk material retains viscoelastic response, with mechanical properties comparable to historical PVC materials. DMA indicated that the material behaves as relatively stiff under short-term loads, but responds as ductile under long-term, low-rate deformations. DFT simulations further support the experimental observations, showing that DOTP is energetically more stable at the PVC surface than within the polymer bulk, and that DOTP self-association is stronger than DOTP–PVC interactions. These findings provide a molecular-level explanation for the observed migration, accumulation, and exudation of the plasticiser.

Overall, this work establishes a mechanistic framework for understanding PVC deterioration in cultural heritage objects. Solid-state NMR emerges as a particularly sensitive technique for the in-depth investigation of ongoing phase separation. The development of non-invasive diagnostic tools inspired by its principles could enable the early identification of PVC heritage assets, and in general, plastic objects, that will be at risk of extensive phase separation (multiple and large *D* within one object), thereby supporting a more targeted preventive conservation strategy for that part of the collection. As a future perspective, extending the DFT approach to other phthalate plasticisers with different molecular sizes and substitution patterns (e.g., *ortho*- and *para*-substituted systems) would provide valuable insights into the relative stability of plasticiser–PVC systems and their susceptibility to phase separation.

**Data Availability**
Most of the data generated or analysed during this study are included in this article and its Supplementary Materials. Additionally, raw datasets are available from the corresponding author on request.